\documentclass[letterpaper]{jpconf}
\usepackage{graphicx}

\begin{document}
\title{SM Higgs Searches at CDF}

\author{B\'arbara \'Alvarez Gonz\'alez}

\address{Universidad de Oviedo}

\begin{abstract}
We present the latest CDF searches for the Standard Model Higgs boson  
with 1.96~TeV center-of-mass energy collisions produced at the Fermilab Tevatron.  The data was 
collected with the CDF II detector at the Tevatron collider 
and correspond to an integrated luminosity from 2 to 4.8~fb$^{-1}$. To achieve  
maximal sensitivity, many channels are analyzed including final states  
from gluon fusion, vector boson fusion, and associated production with  
$W$ and $Z$ bosons.  
\end{abstract}

\section{Introduction}
One of the most fundamental problems in particle physics is 
understanding the mechanism that breaks electroweak symmetry and                                                                               
generates the mass of all known elementary particles. 
The Higgs boson is the only Standard Model (SM) particle that has
not been observed yet and is introduced into the SM theory to explain 
the origin of the electroweak symetry breaking \cite{bib:PeterHiggs}. 
Its mass is a free parameter of the SM, but from direct searches for 
the Higgs boson at the LEP collider the mass of the Higgs boson is 
excluded below 114.4 GeV/c$^2$ at 95\% confidence level (C.L.). 
Indirectly, precision electroweak measurements, performed at LEP and by SLD, CDF, 
and D0, show a preferred low mass Higgs and lead to the 95$\%$ C.L. 
m$_H$ $<$ 157 GeV/c$^{2}$. The 95$\%$ C.L. lower limit obtained from LEP 
is not used in the determination of this limit.

\begin{figure}[h]
  \begin{center}
    \includegraphics[width=0.4\textwidth,clip=]{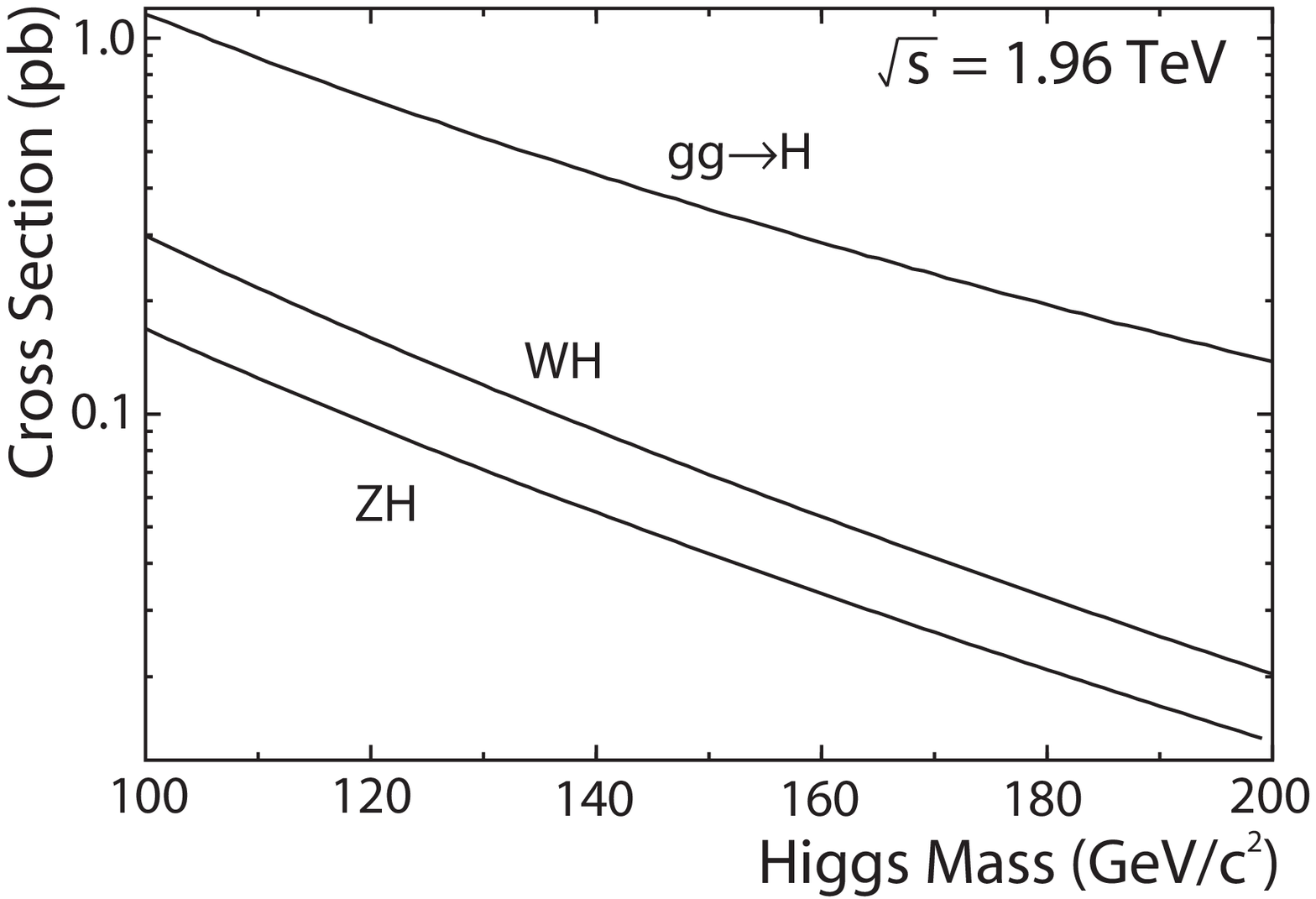} 
    \includegraphics[width=0.28\textwidth,clip=t]{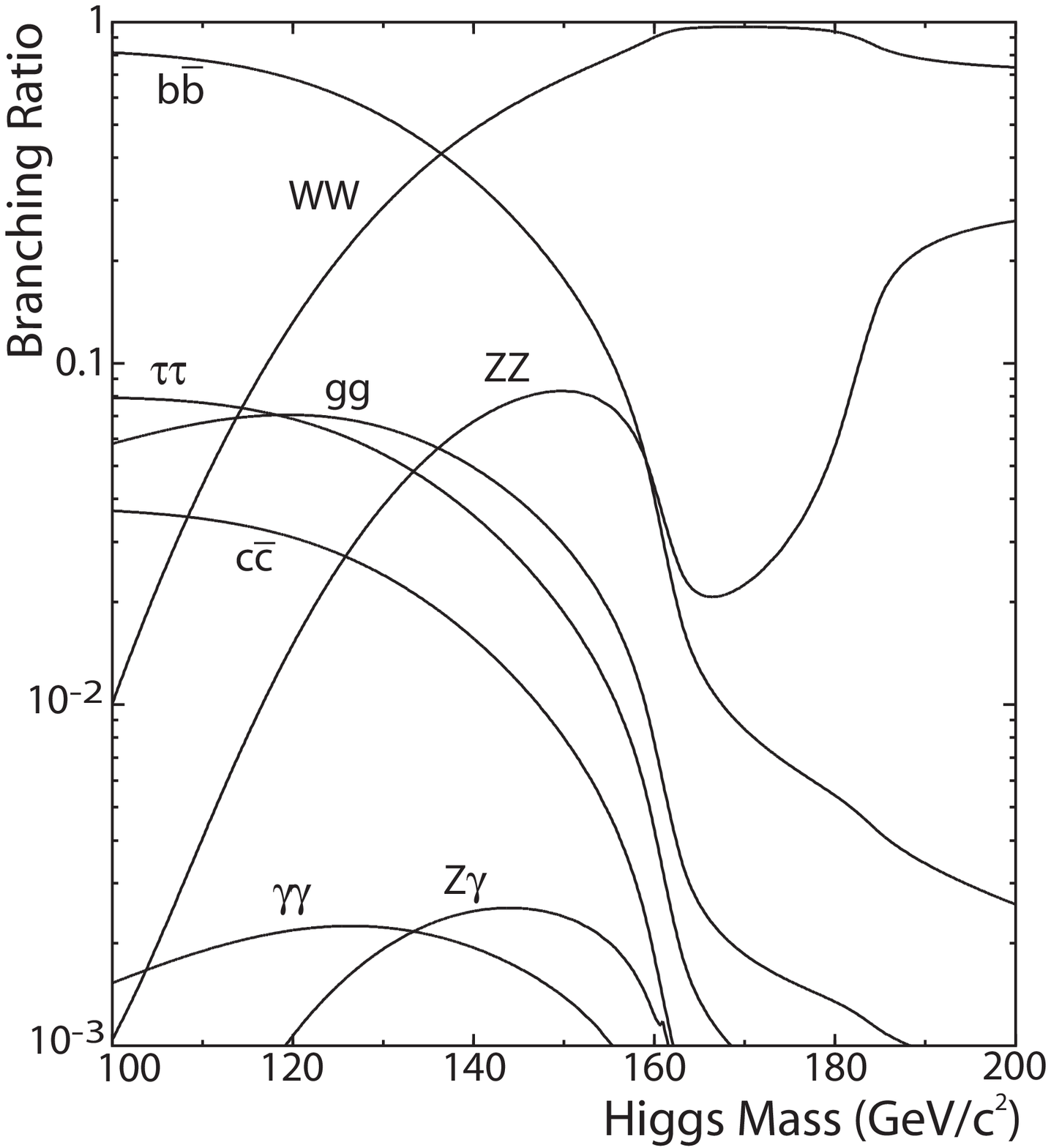}
    \caption[SM Higgs production cross sections for $p\bar{p}$ collisions at 1.96~TeV]
    {SM Higgs production cross sections for $p\bar{p}$ collisions at 1.96~TeV\cite{bib:smxs} (on the left) and 
      branching ratios for the main decays of the SM Higgs boson\cite{bib:sm-mssm-higgsreview} (on the right).
      \label{fig:tevxs}}
  \end{center}
\end{figure}

At the Tevatron, a $p\bar{p}$ collider at Fermilab, the most important SM Higgs 
boson production processes are gluon fusion ($gg \rightarrow H$) and Higgs boson 
production in association with a vector boson ($WH$ or $ZH$)\cite{pp-cross}, 
summarized in Figure~\ref{fig:tevxs} as a function of the Higgs mass ($m_{{H}}$).

The branching ratios for the most relevant decay modes of the SM
Higgs boson are shown in Figure~\ref{fig:tevxs} as functions of $m_{{H}}$.
The search strategy for the Higgs boson at the Tevatron is driven by the decay modes.
For masses below  135~GeV/c$^{2}$, decays to fermion pairs dominate, of which the decay
$H \rightarrow b\bar{b}$ has the largest branching ratio.
Decays to $\tau^+\tau^-$, $c\bar{c}$ and gluon pairs together
contribute less than 15$\%$. In this case, the most promising discovery channels
are $WH$ and $ZH$ with $H \rightarrow b\bar{b}$.
For Higgs boson masses above~135 GeV/c$^{2}$, the $WW$ decay dominates
with an important contribution from $H \rightarrow ZZ$.
Both the direct ($gg \rightarrow H$) and
the associated production ($p\bar{p} \rightarrow WH$ or $ZH$) channels are explored.

\section{SM Higgs Boson Searches at CDF}
An overview of the SM Higgs analyses done at CDF is presented in this section.
The analyses are exploited in two regions: low and high mass region.
At masses below about 135~GeV/c$^{2}$, low mass region, the searches for associated
production,  $p\bar{p} \rightarrow WH,ZH$ are performed in different channels:
$p\bar{p} \rightarrow WH \rightarrow l \nu b\bar{b}$, 
$p\bar{p} \rightarrow ZH \rightarrow \nu \bar{\nu} b\bar{b}$, 
and $p\bar{p} \rightarrow ZH\rightarrow l^{+}l^{-} b\bar{b}$.
At the high mass region, the dominant $ H\rightarrow WW$ decay mode is
best exploited in direct $gg \rightarrow H$ production, however the vector 
boson fusion, and the associated production with $W$ and $Z$ bosons modes 
are also used. 

\subsection{\bf \boldmath $p\bar{p} \rightarrow W H  \rightarrow l \nu b\bar{b}$}
The $W$ decays leptonically, to a lepton 
(electron or muon) and a neutrino, and $ H\rightarrow b\bar{b}$; such searches have
been published by the CDF collaboration on $2.7$~fb$^{-1}$ \cite{bib:cdf-wh-pub}. 
These analyses use very advanced analysis techniques such as neural networks (NNs) or a matrix 
element method to separate a potential signal from the background processes, and also a NN 
is used to separate correctly identified $b$-jets from jets originating from gluons or from light 
or $c$ quarks. The latest results using $4.8$~fb$^{-1}$ of the CDF integrated luminosity, are public 
but have not been published yet~\cite{bib:WH_publicweb}.
The Higgs boson production cross section limits obtained are about three times higher 
than the SM expectation in this channel. The expected and observed limits for a Higgs 
boson mass of 115 GeV/c$^2$ for this channel and for the other two low mass Higgs channels
are shown in Table~\ref{tab:table_limits}, with the corresponding integrated luminosity.

\begin{table}[h]
  \begin{center}
    \begin{tabular}{l|c|c|c}
      \hline
      \hline
      \bf Channel                                                    & \bf Int. Luminosity (fb$^{-1}$) & \bf Expected/SM & \bf Observed/SM  \\
      \hline \hline
      $p\bar{p} \rightarrow W H  \rightarrow l \nu b\bar{b}$         &  4.8                   & 3.8         &  3.3  \\
      \hline
      $p\bar{p} \rightarrow ZH\rightarrow  \nu \bar{\nu} b\bar{b}$   &  3.6                   & 4.2         & 6.1  \\
      \hline 
      $p\bar{p} \rightarrow ZH\rightarrow l^{+}l^{-} b\bar{b}$       &  4.1                   & 6.8         & 5.9  \\
      \hline \hline
    \end{tabular}
  \end{center}
  \caption[Limits]
        {Expected and observed upper limits in Standard Model units 
          for a Higgs boson mass of 115 GeV/c$^2$ for the three low mas 
          Higgs channels with their corresponding integrated luminosities.}
  \label{tab:table_limits}
\end{table}


\subsection{ \bf \boldmath $p\bar{p} \rightarrow ZH\rightarrow  \nu \bar{\nu} b\bar{b}$}
The $Z$ decays into $\nu \bar{\nu}$ and
$ H\rightarrow b\bar{b}$, is also a very sensitive channel, the sensitivity is 
comparable to that obtained in the $WH$ channel. Since the final state is characterized 
by missing transverse energy and two $b$-jets, the QCD multijet processes are the 
dominant background processes. In CDF, a data-driven model is implemented to remove 70\% of the 
QCD background events. The sensitivity of this search is enhanced by $p\bar{p} \rightarrow WH$ events in 
which the charged lepton from the $W$ decay escapes detection; these events have 
the same experimental signature as the $ZH\rightarrow\nu \bar{\nu}$ signal. The 
latest result has been updated using $3.6$~fb$^{-1}$ of the CDF II 
data~\cite{bib:ZH_publicweb}. For a Higgs boson mass of 115 GeV/c$^2$ the expected and observed limits
are given in Table~\ref{tab:table_limits}. 

\subsection{\bf \boldmath $p\bar{p} \rightarrow ZH\rightarrow l^{+}l^{-} b\bar{b}$} 
The $Z$ decays into two charged leptons 
($e^{+}e^{-}$ or $\mu^{+}\mu^{-}$), and the Higgs boson, as well as the other 
two channels, to a $b\bar{b}$ pair. This channel suffers from a smaller $Z$ 
branching fraction, but has lower background, so its sensitivity is not much 
lower than that of the previous two channels. The main background processes 
are $Z$+jets ($b\bar{b}$, and $c\bar{c}$), top pairs, and $ZZ$. Two neural 
networks were implemented, one to improve the energy resolution of the 
two final state jets and the other, a two dimensional NN, to discriminante 
between signal to background events. The latest result presented by the CDF 
collaboration is based on 4.1~fb$^{-1}$ of the CDF integrated luminosity~\cite{bib:cdf-zhllbb}.
The upper limits for a Higgs boson mass of 115 GeV/c$^2$ are shown in Table~\ref{tab:table_limits}.

\subsection{\bf \boldmath $ H\rightarrow WW\rightarrow l^{+}l^{-}\nu \bar{\nu}$}
It is the most sensitive channel for possible SM Higgs bosons with a mass above 
135 GeV/c$^2$. The $WW$ pair issued from a Higgs boson decay has a
spin correlation which is different from the dominant background process, 
electroweak $WW$ production. These spin correlations are transmitted to 
the distributions of observed leptons, providing a handle to separate the 
signal from the background. The invariant mass of the Higgs boson decay
products cannot be reconstructed due to the undetected neutrinos, but
the sensitivity is nevertheless significant.
Other physic processes like Drell-Yan, $W$ boson production associated 
with a photon or jets and top quark pair production are also characterized 
as background processes for this channel.

In order to maximize sensitivity, seven dedicated channels are exploited:
opposite-sign (OS) final states with either zero, one, or two or more jets, 
opposite-sign low dilepton invariant mass (M$_{ll}$), same-sign (SS) dileptons 
and trileptons outside and inside of the $Z$ peak in the final state.
The results from the seven independent channels are combined together to obtain 
the final limits.
Using an integrated luminosity of $4.8~$fb$^{-1}$, an observed (expected) 95\% C.L.
 upper cross section limit of 1.2 (1.2) times the SM prediction for a Higgs mass 
of 165 GeV/c$^2$ is obtained~\cite{bib:cdf-highmass}. 

\section{Conclusions}
\begin{figure}[h]
    \begin{center}
      \includegraphics[width=0.85\textwidth,clip=t]{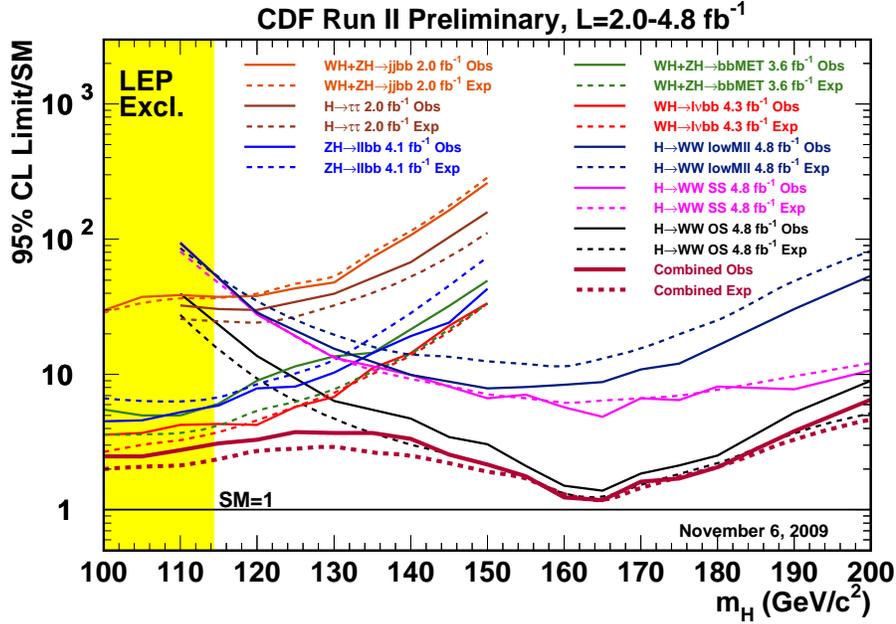}
      \caption[Expected and observed 95\% C.L. upper limits on different channels of Higgs production in CDF]{
    Expected and observed 95\% C.L. upper limits on different channels of Higgs production in CDF,
    expressed as a ratio to the SM cross section expectation times branching. The limits are obtained
    using integrated luminosities from 2 to 4.8 fb$^{-1}$. The dashed line indicates the expected limit and the
    solid line the observed limit. All the separate channel results are combined to obtain the CDF combination (maroon).
        \label{fig:cdf_comb}}
    \end{center}
  \end{figure}

All the presented Standard Model Higgs bosons searches, with other channels 
that are not as sensitive like $H\rightarrow \tau^{+} \tau^{-}$+jets, are 
combined together to extract the maximum sensitivity. No evidence for a Higgs 
boson signal has been reported yet. The CDF Standard Model Higgs boson 
combination results an exclusion limit that is approaching the 
predicted cross sections.
With 2 to 4.8 fb$^{-1}$ of data analyzed at CDF, the 95\% C.L. upper limits 
observed (expected) are factors of 3.1 (2.4) and 1.2 (1.2) higher than the 
Standard Model production cross sections for Higgs boson masses of 115 and 
165 GeV/c$^2$, respectively, shown in Figure~\ref{fig:cdf_comb}~\cite{bib:cdf-comb}. 

The results on direct searches for a SM Higgs boson 
of both Tevatron experiments, CDF and D\O\, are also combined 
to maximize the sensitivity to the Higgs boson~\cite{bib:tev-comb}.

\section{References}
\bibliographystyle{unsrt}
\bibliography{bibliography}

\end{document}